\documentclass[aps,prd,tighten,11pt,showpacs,a4,nofootinbib]{revtex4}

\usepackage{graphicx}
\usepackage{bm}
\usepackage{amsmath,amssymb}
\usepackage{latexsym}
\usepackage{color}

\newcommand{\al}{\alpha}

\newcommand{\bee}{\begin{equation}}
\newcommand{\eee}{\end{equation}}

\begin{document}

\title{Stress-energy tensor of quantized massive fields in static wormhole spacetimes}

\author{Ewa Kocuper, Jerzy Matyjasek\footnote{email: jurek@kft.umcs.lublin.pl} and Kasia Zwierzchowska }

\affiliation{Institute of Physics,
Maria Curie-Sk\l odowska University\\
pl. Marii Curie-Sk\l odowskiej 1,
20-031 Lublin, Poland}

\begin{abstract}
In order to be traversable, the static Lorentzian wormhole must 
be made out of some exotic matter that violates the weak energy 
condition.  The quantized fields are the natural candidates as
their stress-energy tensor, in many cases, possesses desired properties.
In this paper we construct and examine the stress-energy tensor 
of the quantized massive scalar, spinor and vector fields in six static wormhole
spacetimes. We find  that in all considered cases the quantum 
fields violate the Morris-Thorne conditions and do not have the
form necessary to support the wormhole throat. This is in concord
with the previous results and indicates that the massive quantum fields 
make the wormholes less operable.
\end{abstract}
\maketitle

\section{Introduction}

Since their invention the Lorentzian wormholes have been a subject of intensive
studies~\cite{Morris1,Morris2}. Treated as shortcuts in a spacetime 
these strange solutions of the
Einstein field equations opened the discussion of interstellar or even
intergalactic journeys. Moreover, careful analyses of the casual structure
of the spacetimes of the static wormholes indicated the possibility of the
existence of closed timelike curves that lead to the inevitable time-machine
interpretation of such configurations. Surely such strange objects are hard to
construct in practice: it was shown that the stress-energy tensor of the
wormhole has to satisfy the more or less  exotic conditions, excluding the 
possibility
of constructing an operating wormhole out of any known form of the ordinary
classical matter. On the other hand however, it is a well-known fact that
the quantized fields may violate classical energy conditions and therefore
may appear to be of sufficient exoticity to maintain a wormhole.

Having an operating traversable wormhole may be a source of problems. Indeed,
it would  not only be an interesting and perhaps useful device for us, but also
it could play a similar role for some advanced civilization. Following this line 
of reasoning one 
can regard the fundamental obstacles one encounters in projecting and 
constructing
wormholes as rather fortunate. 

To be a good wormhole the solution of the Einstein field equations
must satisfy some quite restrictive requirements, such as the existence of a 
throat 
that connects two spacetime  regions, the nonexistence of the event horizon, 
traversability by humans (that sets the scale for the curvature) and 
perturbative 
stability. There have been numerous attempts to construct the exact solutions of 
the
Einstein field equations describing a traversable wormhole. Various features
of such solutions are well documented and have been analyzed in numerous 
papers, 
e.g., in Refs.~\cite{VisserBook,Lobo,KirillBook,ayon}
(and the references cited therein). On the other hand however, less is known 
about 
wormholes generated or influenced
by the quantized fields. This is perhaps not particularly strange as the
calculations of this sort are very complicated and have to be carried out 
with the aid of numerical methods. 
On the other hand however, one may, as a preliminary step, construct and
analyze the renormalized stress-energy tensor of the quantized fields of
various types in a spacetime of interest, and decide if the obtained $\langle
T^{b}_{a}\rangle$ has the appropriate form to support a
traversable wormhole spacetime. Such calculations have been initiated by Taylor 
et al.
in Ref~\cite{TaylorPaul}, where the stress-energy tensor of the quantized 
massive scalar field
has been constructed and analyzed  for the five types of wormhole 
spacetimes (types A - E of Sect.~\ref{main}). 
They used the general formula constructed in Ref.~\cite{Paul} that is
valid for an arbitrary static and spherically symmetric spacetime and concluded
that for both minimally and conformally coupled massive scalar fields the
stress-energy tensor does not have the form needed to support the wormhole
geometry and that the exotic conditions were satisfied for clearly
unphysical values of the coupling constant. Other results, especially 
those obtained in 
Refs.~\cite{Carlson,Khatsymovsky,Popov2014,Bezerra,PopovS2,Popov2001,Khu,
Garattini} 
were devoted to various aspects of the quantum field theory in the wormhole 
spacetimes.
In this regard the (numerical) construction of the semiclassical wormhole
reported in Ref.~\cite{PopovSushkov} is of principal interest.

The analytic approximation used in Ref.~\cite{Paul} is
based on the WKB approximation of the solutions of the massive scalar field
equation in the spherically symmetric spacetime and the summation of the thus 
obtained modes by means of the Abel-Plana
method. In the large-mass limit it is equivalent to the Schwinger-DeWitt 
expansion based on the heat kernel 
expansion~\cite{Bryce1,IvanBook,Frolov,Kocio1,Kocio2}; 
to obtain the
lowest (i. e. $m^{-2}$) terms, one has to use the sixth-order WKB approximation.
The same result can be obtained within the framework of the Schwinger-DeWitt 
approach. Now one has to retain the coincidence limit of the first nontrivial  
Hadamard-DeWitt coefficient, $a_{3}(x,x').$
Numerical calculations reported in Ref.~\cite{Paul1} indicated that the
Schwinger-DeWitt method provides a good approximation of the renormalized
stress-energy tensor of the massive scalar field with an arbitrary curvature
coupling as long as the mass of the field remains sufficiently large.

Some time ago we calculated the approximate renormalized $\langle
T^{b}_{a}\rangle$ of the massive scalar, spinor and vector fields
in an arbitrary spacetime functionally differentiating the effective action
(constructed from the coincidence limit of
Hadamard-DeWitt coefficient $a_{3}$) with respect to
the metric tensor~\cite{Kocio1,Kocio2}. (The next-to-leading term calculated 
from the coefficient $a_{4}$ has been 
presented in Ref.~\cite{next} and the generalization of the method to the 
higher dimensions has been discussed in 
Refs.~\cite{kocio_dim_a,kocio_dim_b,kocio_dim_c}.)
The resulting stress-energy tensor is rather involved
and consists of a few dozen  terms, such as terms cubic in curvature or
involving its fourth covariant derivatives. The obtained $\langle
T^{b}_{a}\rangle$ generalizes the results derived in Ref.~\cite{Frolov} by 
Frolov and
Zel'nikov for vacuum type-D geometries. It is expected that for sufficiently 
massive
fields, i.e., where the Compton length is much smaller than the
characteristic radius of curvature, the approximation is quite reasonable. The 
usual
argument in this regard is that in such a case the nonlocal contribution to
the total effective action may be neglected. 

In this paper we shall extend the analyses of Ref.~\cite{TaylorPaul} to the case 
of the
massive spinor and  vector fields and construct the renormalized stress-energy 
tensor in the six models of the traversable wormholes. The stress-energy tensor 
of the 
quantized massive scalar field in the first five configurations has been 
constructed
earlier. We believe our spinor and vector results as well as the result
obtained for the wormhole supported by the phantom energy are new.
Although the stress-energy tensor of the quantized fields constructed in the 
wormhole spacetime 
is of great interest and importance  in its own right, 	 
we will extend our discussion to the  question of whether the quantized matter 
is of
sufficient exoticity to support the traversable wormhole.  
Ideally, the problem should be treated within the framework of the semiclassical
gravity and it seems that the stress-energy tensor constructed within the 
framework
of the Schwinger-DeWitt approximation is very well suited for this. Indeed,
it depends functionally on a general metric and can be used in virtually any
spacetime provided the above-mentioned conditions are satisfied. 
This, however, would require the numerical analysis of the complex system of the 
sixth-order
differential equations in the spirit of Ref.~\cite{PopovSushkov} that is beyond 
the scope 
of the present paper. Instead, we will confine ourselves to the analysis of the 
properties 
of the stress-energy tensor at the throat and its closest vicinity.  
The general expressions describing the stress-energy tensor are usually too 
long, complicated
and not particularly illuminating and will be not  presented here.\footnote{The 
general expressions 
describing the stress-energy tensor of the quantized massive 
scalar, spinor and vector fields can be obtained on request
from the second author.}

The paper is organized as follows. In Sec.~\ref{mt} the general Morris-Thorne 
conditions are 
introduced. In Sec.~\ref{secT} we discuss the general stress-energy tensor of 
the quantized 
massive fields in the large-mass limit. The analysis of the stress-energy 
tensor and the
Morris-Thorne conditions in the six exemplary wormhole spacetimes  (cases A - F) 
are presented 
in Sec.~\ref{main}. Finally,  Sec.~\ref{fin} contains summary and a brief 
discussion of possible
extensions and generalizations of the presented results.

\section{The general Morris-Thorne wormhole}
\label{mt}

The line element of the static and spherically symmetric wormhole is given by
\begin{equation}
ds^{2}=-\exp \left( 2\phi \right) dt^{2}+\left( 1-\frac{b}{r}\right)
^{-1}dr^{2}+r^{2}\left( d\theta ^{2}+\sin^{2} \theta d\varphi ^{2}\right) ,
\label{gen_el}
\end{equation}
where $\phi(r)$ and $b(r)$ are the redshift and the shape functions, 
respectively. 
The radial coordinate decreases from infinity to its minimal value $r=r_{0}$ 
(describing
the throat) and subsequently it goes to infinity.

Simple calculations show that in the static orthonormal frame the Einstein field 
equations
can be written in the form
\begin{equation}
 \rho = T_{\hat{t}\hat{t}} = \frac{1}{8\pi r^{2}} \frac{d}{dr}b
\end{equation}
and 
\begin{equation}
 \tau =  -T_{\hat{r}\hat{r}} = \frac{1}{8 \pi}\left[ \frac{b}{r^{3}} 
 - \frac{2}{r} \left( 1-\frac{b}{r}\right) \frac{d}{dr} \phi\right].
\end{equation}
The equations for the mass-energy, $\rho,$ and the radial tension, $\tau,$
are supplemented by the  third equation 
\begin{equation}
 p = \frac{r}{2}\left[( \rho-\tau) \frac{d}{dr}\phi-\frac{d}{dr} \tau \right]
\end{equation}
for the lateral pressure $p = T_{\hat{\varphi}\hat{\varphi}} = 
T_{\hat{\theta}\hat{\theta}}.$
The traversable wormhole configurations require satisfying at the throat
(located at $r=r_{0}$) the  following minimal set of  conditions 
\begin{equation}
 \tau_{0} >0
 \label{var1}
\end{equation}
and 
\begin{equation}
 \frac{\tau_{0} -\rho_{0}}{|\rho_{0}|} \geq 0.
 \label{var2}
\end{equation}
The first condition follows (via the Einstein field equations)
from the finiteness of $\rho_{0}$ and from the
absence of the event horizon.
The second inequality is the flaring-out condition~\cite{Morris1}.

Now, let us assume that we have the solution of the Einstein field 
equations describing an operating wormhole and consider quantum fields
in the spirit of the quantum field theory in curved background,
i.e., the mean value of the stress-energy tensor of the quantum fields
does not contribute to the source term. In general, depending on its 
characteristic, 
the action of the quantum field can be either destructive or constructive.
Indeed, if the Morris-Thorne conditions are not satisfied, the quantum effects 
tend to destroy
or at least make the wormhole less functional. In the next section
we shall construct the stress-energy tensors of the massive scalar, spinor
and vector fields in the spacetimes of the six wormhole configurations.

\section{Stress-energy tensor of the massive scalar, spinor and vector fields}
\label{secT}

The approximate one-loop effective action of the quantized massive scalar, 
spinor 
and vector fields  is given by
\begin{eqnarray}
W^{(1)}\,&=&\,{\frac{1}{192 \pi^{2} m_{\star}^{2}}} \int d^{4}x g^{1/2} \left[
\alpha_{1} R\Box R\,
+\,\alpha_{2} R_{p q} \Box R^{p q}\,
+\, \alpha_{3} R^{3} \right.  \nonumber \\
&+& \alpha_{4} R R_{p q } R^{p q} 
+\,\alpha_{5} R R_{p q a b} R^{p q a b}\,
+\, \alpha_{6} R^{p}_{q} R^{q}_{a} R^{a}_{p} 
+ \alpha_{7}R^{p q} R_{a b} R^{a ~ b}_{~ p ~ q}  \nonumber \\
&+&\left. \, 
\,\alpha_{8} R_{p q} R_{~ c a b}^{p} R^{q c a b}\,
+\, \alpha_{9} {R_{a b}}^{p q} {R_{p q}}^{c d} {R_{ c d}}^{a b} 
+ \alpha_{10} R^{a ~ b}_{~ p ~ q} R^{p ~ q}_{~ c ~ d} R^{c ~d}_{~ a ~ b}\right]  
\nonumber \\
&=&{\frac{1}{192 \pi^{2} m_{\star}^{2}}}\sum_{i = 1}^{10} \alpha_{i} W_{i},
\label{action}
\end{eqnarray}
where $m_{\star}$ is the effective mass and
\begin{equation}
 \alpha_{i} = \frac{\alpha_{i}^{(0)}}{n_{0}} + 
\frac{\alpha_{i}^{(1/2)}}{n_{1/2}} + \frac{\alpha_{i}^{(1)}}{n_{1}}.
\end{equation}
The coefficients $\alpha_{i}^{(s)}$ are tabulated in Table I and the parameters 
$n_{s}$ 
are defined by the simple relations
$m_{0}^{2} = n_{0} m_{\star}^{2}, $
$ m_{1/2}^{2} = n_{1/2} m_{\star}^{2}$
and 
$ m_{1}^{2} = n_{1} m_{\star}^{2}.$
\begin{table}[h]
\caption{The coefficients $\al_{i}^{(s)}$ for the massive scalar,  spinor, and 
vector fields}
\begin{tabular}{|c|c|c|c|}
\toprule
&s=0&s=1/2&s=1\\ \hline
$\al^{(s)}_{1}$&$\frac{1}{2}\xi^{2}\,-\,\frac{1}{5}\xi$\,+\,$\frac{1}{56}
$&$-\frac{3}{140}$&$-\frac{27}{280}$\\	
$\al^{(s)}_{2}$&$\frac{1}{140}$&$\frac{1}{14}$&$\frac{9}{28}$\\$\al^{(s)}_{3}
$&$\left(\frac{1}{6}-\xi\right)^{3}$&$\frac{1}{432}$&$-\frac{5}{72}$\\
$\al^{(s)}_{4}$&$-\frac{1}{30}\left(\frac{1}{6}-\xi\right)$&$-\frac{1}{90}
$&$\frac{31}{60}$\\$\al^{(s)}_{5}$&$\frac{1}{30}\left(\frac{1}{6}-\xi\right)$&$
-\frac{7}{720}$&$-\frac{1}{10}$\\$\al^{(s)}_{6}$&$-\frac{8}{945}$&$-\frac{25}{
378}$&$-\frac{52}{63}$\\$\al^{(s)}_{7}$&$\frac{2}{315}$&$\frac{47}{630}$&$-\frac
{19}{105}$\\	
$\al^{(s)}_{8}$&$\frac{1}{1260}$&$\frac{19}{630}$&$\frac{61}{140}$\\	
$\al^{(s)}_{9}$&$\frac{17}{7560}$&$\frac{29}{3780}$&$-\frac{67}{2520}$\\	
$\al^{(s)}_{10}$&$-\frac{1}{270}$&$-\frac{1}{54}$&$\frac{1}{18}$\\ \botrule	
\end{tabular}\label{table1}\end{table}
The $s$-spin field can be excluded from the calculations by taking the limit 
$n_{s} \to \infty.$ Without loss of generality the parameter $n_{0}$ can be 
normalized by taking 
$m_{\star}$ to be the effective mass of the scalar fields, i.e.,
$n_{0} =1$ if the massive scalar field is present and $n_{0} =\infty$ if it is 
absent.

The renormalized stress-energy tensor is given by 
\begin{equation}
\langle T^{a b}\rangle  =
{\frac{2}{g^{1/2}}}{\frac{\delta }{\delta g_{ab}}}W_{ren} = \frac{1}{96\pi^{2} 
m_{\star}^{2}{g^{1/2}}}    
\sum_{i=1}^{10} \alpha_{i} \frac{\delta}{\delta g_{ab}} W_{i}
\end{equation}
and the full form of the functional derivatives of $W_{i}$ with respect to the 
metric tensor was given in Refs.~\cite{Kocio1,Kocio2}.
It is convenient to rewrite the stress-energy tensor in the following form
\begin{equation}
 \langle T_{a}^{b} \rangle = \frac{1}{96 \pi^{2} m^{2}_{\star}} 
\left(\frac{t_{a}^{b}}{n_{0}} 
 + \frac{s_{a}^{b}}{n_{1/2}} + \frac{v_{a}^{b}}{n_{1}} \right),
 \label{strr}
\end{equation}
where $t_{a}^{b},$ $s_{a}^{b}$ and $v_{a}^{b}$ are the purely geometric parts of 
the stress-energy tensor of 
the scalar, spinor and vector fields, respectively.

Using ordinary tensor components rather than proper reference frame components 
one has 
\begin{equation}
 \rho=  - \langle T_{t}^{t}\rangle,
\end{equation}
\begin{equation}
 \tau =  - \langle T_{r}^{r}\rangle
\end{equation}
and
\begin{equation}
 p =  \langle T_{\theta}^{\theta}\rangle =  \langle T_{\phi}^{\phi}\rangle.
\end{equation}
Our analysis will be confined to the closest vicinity of the throat.  
We will concentrate on the scalar, vector and spinor quantum fields propagating 
in the five wormhole 
spacetimes (cases A - E) considered earlier by Taylor et al. Additionally, we 
will also consider the case of the wormhole
supported by phantom energy~\cite{Oleg}. It should be noted that for the scalar 
fields 
with arbitrary curvature our approach gives precisely the same results as these 
presented 
in Ref.~\cite{TaylorPaul} and for the reader's convenience we shall also display 
them in this paper. 
We only remark that taking into account fundamental differences
of the two methods the equality of the results, although expected, is indeed 
really impressive.

We conclude this section with a few words on implementation. The functional 
derivatives of the
effective action (\ref{action}) with respect to the general metric tensor as 
well as basic simplifications
of the thus obtained result have been constructed with the aid of the 
FORM~\cite{Jos1,Jos2}. The result
has been converted into the Mathematica and Maple syntaxes for further tensor 
simplifications
and for  calculations in the local maps. The final result has also been derived 
 
from the effective action constructed from the full form of the coefficient 
$[a_{3}].$
In both cases the FORM calculations have been successfully executed within a 
fraction of a second.

\section{Quantum fields and the Morris-Thorne conditions}
\label{main}

Now we shall construct the stress-energy tensor of the quantized fields in a 
large-mass limit in six
wormhole spacetimes and check if it satisfies the conditions (\ref{var1}) and 
(\ref{var2}). In general, the stress-energy 
tensor for the spacetime described by the line element (\ref{gen_el}) is quite 
complicated and to 
avoid  the presentation of the unnecessary complex formulas  we will focus on 
its value at the throat.
We shall start our survey with the simplest case of the zero-tidal wormhole.

\subsection{Zero-tidal wormhole} 
The simplest traversable wormhole with the zero radial tides is described 
by the line element (\ref{gen_el}) with 
\begin{equation}
\phi \left( r\right) =0,\qquad b(r)=r_{0}.
\end{equation}
Because of its  form one expects massive simplifications in the stress-energy 
tensor. 
Indeed, for the scalar, spinor and vector fields, after some  algebra, one has
\begin{equation}
 \langle T_{t}^{t}\rangle = \frac{1}{96 \pi^{2}m_{\star}^{2} 
r_{0}^{6}}\left(\frac{43-252 \xi }{560 n_{0}} -\frac{3}{56 
n_{1/2}}-\frac{53}{560 n_{1}}   \right),
 \label{ttA}
\end{equation}
\begin{equation}
 \langle T_{r}^{r}\rangle = \frac{1}{96 \pi^{2}m_{\star}^{2} 
r_{0}^{6}}\left(\frac{23-84 \xi }{560 n_{0}} + \frac{3}{70 n_{1/2}} 
+\frac{83}{560 n_{1}}  \right)
 \label{rrA}
\end{equation}
and
\begin{equation}
 \langle T_{\theta}^{\theta}\rangle = \frac{1}{96 \pi^{2}m_{\star}^{2} 
r_{0}^{6}}\left( +\frac{5-21 \xi }{56 n_{0}} + \frac{33}{560 
n_{1/2}}+\frac{19}{280 n_{1}}     \right).
 \label{angA}
\end{equation}
Although the maximal power of $\xi$ is 3 (as can be seen from 
Table~\ref{table1}) 
the stress-energy tensor is linear in $\xi$. It is because the curvature scalar 
vanishes,
and, consequently, the first and  third terms of~(\ref{action}) do not 
contribute to the final result.

The Morris-Thorne conditions require the tension at the throat to be positive, 
i.e., $\tau_{0}>0.$ Moreover, in order to
construct the traversable wormhole  the tension should be greater than or equal 
to the energy density, $\tau_{0} \geq \rho_{0}.$ 
Inspection of the stress-energy tensor  shows that the tension of the spinor and 
vector fields is negative whereas the
energy density is positive. It follows then that both conditions are violated 
and the spinor and vector fields tend to
destroy the wormhole. The analysis carried out in Ref.~\cite{TaylorPaul} showed
that the Morris-Thorne conditions cannot be satisfied
simultaneously by the scalar field. Indeed, from Eqs. (\ref{ttA}) and 
(\ref{rrA}) one concludes that the tension is positive for $\xi > 23/84$ and 
the 
condition (\ref{var2}) is satisfied for $\xi \leq 5/42.$

\subsection{The simple wormhole}
As a second example let us consider a slightly more complicated
ultrastatic wormhole with
\begin{equation}
\phi \left( r\right) =0,\qquad b\left( r\right) =\frac{r_{0}^{2}}{r}.
\end{equation}
Being still relatively simple, the stress-energy tensor at the throat can be 
written in the following compact form
\begin{equation}
  \langle T_{t}^{t}\rangle = -\frac{1}{96 \pi^{2}m_{\star}^{2} r_{0}^{6}}\left(  
\frac{19320 \xi ^3-33180 \xi ^2+11242 \xi -1033}{210 n_{0}}+ \frac{19}{21 
n_{1/2}}+ \frac{381}{70 n_{1}}\right),
\end{equation}
\begin{equation}
  \langle T_{r}^{r}\rangle = -\frac{1}{96 \pi^{2}m_{\star}^{2} r_{0}^{6}}\left(  
\frac{4200 \xi ^3-5460 \xi ^2+1806 \xi -187}{210 n_{0}} - \frac{23}{105 
n_{1/2}}-\frac{89}{70 n_{1}}          \right)
\end{equation}
and
\begin{equation}
  \langle T_{\theta}^{\theta}\rangle = -\frac{1}{96 \pi^{2}m_{\star}^{2} 
r_{0}^{6}}\left(\frac{19320 \xi ^3-33180 \xi ^2+11634 \xi -1213}{210 n_{0}} 
  - \frac{131}{105 n_{1/2}} -\frac{527}{70 n_{1}}  \right).
\end{equation}
The energy density of the massive scalar field is positive for $0.1622 \leq \xi 
\leq 0.2532 $ and for $\xi \geq 1.302$
whereas the radial tension is negative for $\xi < 0.8604.$ Finally, the lateral 
pressure is
positive for $\xi\leq 1.2876.$
Now, let us return to the Morris-Thorne conditions and consider the spinor and 
vector fields first. The  condition (\ref{var1}) is violated
as the radial tension in both cases is negative. Similarly, since  
$\rho_{0}$ is positive, the second condition is also violated. The case of the 
scalar 
fields is slightly more complicated: simple manipulations
indicate that the Morris-Thorne conditions are satisfied for (approximately) 
$0.8604 \leq \xi \leq 1.4222.$

\subsection{The ``absurdly benign'' wormhole}

The wormhole configurations considered in this subsection are characterized by
the locality of the distribution of the exotic matter. The matter is confined 
to an arbitrary small region in the closest vicinity of the throat, i.e.,
\begin{equation}
\phi \left( r\right) =0,\qquad b\left( r\right) =\frac{r_{0}\left(
a+r_{0}-r\right) ^{2}}{a^{2}}
\end{equation}
for $r_{0}\leq r <r_{0}+a$ and $b(r)= 0$ outside this region.

Adopting the notation of Sec.~\ref{secT} (Eq.~\ref{strr}) the (rescaled)
time component of the stress-energy tensor in each of the considered cases is 
given by
\begin{eqnarray}
t_{t}^{t} &=& \frac{\xi^3 \left(-161280 a^3 r_{0}^2-349440 a^2 r_{0}^3-161280 a 
r_{0}^4\right)}{1680 a^5 r_{0}^6}\nonumber \\
&+&\frac{\xi^2 \left(23520 a^4 r_{0}+191520 a^3 r_{0}^2+342720 a^2 
r_{0}^3+164640 a r_{0}^4+6720 r_{0}^5\right)}{1680 a^5 r_{0}^6}\nonumber \\
&+&\frac{\xi  \left(-756 a^5-11536 a^4 r_{0}-59080 a^3 r_{0}^2-95200 a^2 
r_{0}^3-46368 a r_{0}^4-2688 r_{0}^5\right)}{1680 a^5r_{0}^6}\nonumber \\
&+&\frac{129 a^5+1202 a^4 r_{0}+5134 a^3 r_{0}^2+7856 a^2 r_{0}^3+3884 a 
r_{0}^4+264 r_{0}^5}{1680 a^5 r_{0}^6},
\end{eqnarray}
\begin{equation}
 s_{t}^{t} =-\frac{45 a^5+334 a^4 r_{0}+668 a^3 r_{0}^2+632 a^2 r_{0}^3+230 a 
r_{0}^4+12 r_{0}^5}{840 a^5 r_{0}^6},
\end{equation}
and
\begin{equation}
 v_{t}^{t} = -\frac{159 a^5+2302 a^4 r_{0}+7026 a^3 r_{0}^2+9584 a^2 
r_{0}^3+3972 a r_{0}^4+216 r_{0}^5}{1680 a^5 r_{0}^6}.
\end{equation}
Similarly, for the radial and angular components one has 
\begin{eqnarray}
t_{r}^{r} &=&\frac{\xi^3 \left(-80640 a^2 r_{0}^2-107520 a r_{0}^3\right)}{1680 
a^4 r_{0}^6}\nonumber \\
&+&\frac{\xi ^2 \left(6720 a^3 r_{0}+70560 a^2 r_{0}^2+94080 a r_{0}^3+13440 
r_{0}^4\right)}{1680 a^4r_{0}^6}\nonumber \\
&+&\frac{\xi  \left(-252 a^4-3360 a^3 r_{0}-19824 a^2 r_{0}^2-25536 a 
r_{0}^3-5376 r_{0}^4\right)}{1680 a^4 r_{0}^6}\nonumber \\
&+&\frac{69 a^4+450 a^3 r_{0}+1878 a^2 r_{0}^2+2272 a r_{0}^3+552 r_{0}^4}{1680 
a^4 r_{0}^6},
\end{eqnarray}
\begin{equation}
s_{r}^{r} =\frac{36 a^4+96 a^3 r_{0}+201 a^2 r_{0}^2+164 a r_{0}^3+36 
r_{0}^4}{840 a^4 r_{0}^6},
\end{equation}
\begin{equation}
v_{r}^{r} = \frac{83 a^4+254 a^3 r_{0}+842 a^2 r_{0}^2+928 a r_{0}^3+216 
r_{0}^4}{560 a^4 r_{0}^6}
\end{equation}
and
\begin{eqnarray}
t_{\theta}^{\theta} &=&\frac{\xi^3 \left(-60480 a^3 r_{0}^2-215040 a^2 
r_{0}^3-80640 a r_{0}^4\right)}{840 a^5 r_{0}^6}\nonumber \\
&+&\frac{\xi^2 \left(10080 a^4 r_{0}+84840 a^3 r_{0}^2+198240 a^2 r_{0}^3+85680 
a r_{0}^4+3360 r_{0}^5\right)}{840 a^5 r_{0}^6}\nonumber \\
&+&\frac{\xi  \left(-315 a^5-5838 a^4 r_{0}-29484 a^3 r_{0}^2-55776 a^2 
r_{0}^3-25200 a r_{0}^4-1344 r_{0}^5\right)}{840 a^5 r_{0}^6}\nonumber \\
&+&\frac{75 a^5+876 a^4 r_{0}+3198 a^3 r_{0}^2+5132 a^2 r_{0}^3+2325 a 
r_{0}^4+138 r_{0}^5}{840 a^5 r_{0}^6},
\end{eqnarray}
\begin{equation}
s_{\theta}^{\theta} =
\frac{99 a^5+906 a^4 r_{0}+1803 a^3 r_{0}^2+1732 a^2 r_{0}^3+654 a r_{0}^4+36 
r_{0}^5}{1680 a^5 r_{0}^6},
\end{equation}
\begin{equation}
v_{\theta}^{\theta} =\frac{19 a^5+512 a^4 r_{0}+1546 a^3 r_{0}^2+2164 a^2 
r_{0}^3+939 a r_{0}^4+54 r_{0}^5}{280 a^5 r_{0}^6}.
\end{equation}
%
In this case the stress-energy tensor of the quantized massive spinor and vector 
fields does not satisfy the Morris-Thorne 
conditions as the energy density (and the lateral pressure) treated as a 
function of $a/r_{0}$
is always positive whereas the radial tension is always negative.
On  the other hand, for the massive scalar fields there is a region in the 
parameter space
where the conditions are satisfied. However, it is possible only for very exotic 
and apparently unphysical values of the 
curvature coupling constant.

\subsection{Wormhole with finite radial cutoff}

This is the class of the wormhole geometries in which the zero-tidal-force 
solution is joined with the Schwarzschild solution at some finite radius, $a$.
The line element in the vicinity of the throat $(r_{0}\leq r \leq a)$ 
is
\begin{equation}
\phi \left( r\right) =0,\qquad b\left( r\right) =r_{0}\left( \frac{r}{r_{0}}
\right) ^{\left( 1-\eta \right) },
\label{dtt}
\end{equation}
where $0<\eta <1.$
The stress-energy tensor at the throat is given by
\begin{eqnarray}
t_{t}^{t} &=& \frac{132 \eta^5+403 \eta^4-92 \eta^3-490 \eta^2+112 \eta 
+64}{1680 r_{0}^6}\nonumber \\
&-&\frac{4 (\eta -1)^2 \left(3 \eta ^2+5 \eta +1\right) \xi ^3}{ 
r_{0}^6}\nonumber \\
&+&\frac{\left(\eta-1 \right)\left(2 \eta^4+ 13 \eta ^3+10 \eta^2-18 \eta -2 
\right) \xi ^2}{ r_{0}^6} \nonumber \\
&-&\frac{\left(48 \eta^5+177 \eta^4-41 \eta^3-245 \eta ^2+64 \eta +24\right) \xi 
}{60 r_{0}^6},
\end{eqnarray}
\begin{equation}
 s_{t}^{t} =  -\frac{6 \eta^5+25 \eta^4+20 \eta^3+14 \eta -20}{840 r_{0}^6}
\end{equation}
and
\begin{equation}
 v_{t}^{t} = -\frac{36 \eta^5+143 \eta^4+36 \eta^3-210 \eta ^2+112 \eta -64}{560 
r_{0}^6}.
\end{equation}
For the radial component of the stress-energy tensor one has
\begin{eqnarray}
 t_{r}^{r} &=& \frac{69 \eta^4+146 \eta^3-210 \eta^2+64}{1680 r_{0}^6} \nonumber 
\\
&-&\frac{4 (\eta -1)^2 (2\eta +1) \xi^3}{r_{0}^6} \nonumber \\
&+&\frac{\left(\eta^4+5 \eta^3-8 \eta^2+2\right) \xi^2}{r_{0}^6} \nonumber \\
&-&\frac{\left(8 \eta^4+22 \eta^3-35 \eta^2+8\right) \xi }{20r_{0}^6},
\end{eqnarray}
\begin{equation}
 s_{r}^{r} = \frac{9 \eta^4+23 \eta^3+40}{1680 r_{0}^6},
\end{equation}
\begin{equation}
 v_{r}^{r} = \frac{27 \eta^4+62 \eta^3-70 \eta ^2+64}{560 r_{0}^6}.
\end{equation} 
Finally,  the angular component is given by
\begin{eqnarray}
t_{\theta}^{\theta} &=& \frac{138 \eta^5+495 \eta ^4-19 \eta^3-840 \eta^2+504 
\eta -128}{1680r_{0}^6}\nonumber \\
&-&\frac{2 (\eta -1)^2 \left(6 \eta ^2+13 \eta -4\right) 
\xi^3}{r_{0}^6}\nonumber\\
&+&\frac{\left(4 \eta^5+23 \eta^4-\eta^3-64 \eta^2+46 \eta-8\right) \xi^2}{2 
r_{0}^6}\nonumber \\
&-&\frac{\left(32 \eta^5+130 \eta^4-11 \eta^3-280\eta^2+176 \eta -32\right) \xi 
}{40 r_{0}^6},
\end{eqnarray}
\begin{equation}
 s_{\theta}^{\theta} =\frac{36 \eta^5+141 \eta^4+97 \eta ^3+84 \eta -160}{3360 
r_{0}^6}
\end{equation}
and
\begin{equation}
 v_{\theta}^{\theta} = \frac{54 \eta^5+201 \eta^4+23 \eta^3-280 \eta^2+168 \eta 
-128}{560 r_{0}^6}.
 \label{dang}
\end{equation}

Inspection of the formulas (\ref{dtt}-\ref{dang}) shows that  the spinor and 
vector  tend to destroy the wormhole.
The radial tension is always negative as there are no roots in the interval $0 
\leq \eta \leq 1.$
Similarly, although in the case of the scalar field the parameter space is 
two-dimensional 
there are no configurations that support wormholes. 

\subsection{The proximal Schwarzschild wormhole}

Now, let us consider spacetime that is  similar to the Schwarzschild solution
with one notable modification: the $g_{tt}$ component of the metric tensor
is given by
\begin{equation}
 g_{tt} = - \left( 1-\frac{r_{0}}{r} + \frac{\varepsilon}{r^{2}}\right),
\end{equation}
 $b(r) = r_{0}$
and the line element takes the form
\begin{equation}
ds^{2}=-\left( 1-\frac{r_{0}}{r}+\frac{\varepsilon}{r^{2}}\right)
dt^{2}+\left( 1-\frac{r_{0}}{r}\right) ^{-1}dr^{2}+r^{2}\left( d\theta ^{2}+\sin
^{2}\theta d\varphi ^{2}\right) .
\end{equation}
The  role of the small positive parameter $\varepsilon$ is to prevent 
the existence of the event horizon
and to make the wormhole traversable. Now, making use of  the general formulas 
of Refs.~\cite{Kocio1,Kocio2}
one gets
\begin{eqnarray}
t_{t}^{t} &=&-\frac{\xi^3 \left(13 r_{0}^6-54 r_{0}^4 \epsilon +24 r_{0}^2 
\epsilon ^2+64 \epsilon^3\right)}{8 r_{0}^6 \epsilon^3}\nonumber \\
&+&\frac{\xi^2 \left(47r_{0}^6-150 r_{0}^4 \epsilon -22 r_{0}^2 \epsilon ^2-68 
\epsilon^3\right)}{16 r_{0}^6 \epsilon^3}\nonumber \\
&+&\frac{\xi  \left(-243 r_{0}^6+738r_{0}^4 \epsilon +238 r_{0}^2 \epsilon 
^2+400 \epsilon^3\right)}{240 r_{0}^6 \epsilon^3}\nonumber \\
&-&\frac{-358 r_{0}^6+1044 r_{0}^4 \epsilon +491r_{0}^2 \epsilon ^2+442 
\epsilon^3}{3360 r_{0}^6 \epsilon^3},
\end{eqnarray}
\begin{equation}
s_{t}^{t} = -\frac{-269 r_{0}^6+486 r_{0}^4 \epsilon +822 r_{0}^2 \epsilon 
^2+1396 \epsilon ^3}{6720 r_{0}^6 \epsilon ^3}
\end{equation}
and
\begin{equation}
v_{t}^{t}=-\frac{-1074 r_{0}^6+1620 r_{0}^4 \epsilon +3223 r_{0}^2 \epsilon 
^2+7738 \epsilon ^3}{3360 r_{0}^6 \epsilon ^3}.
\end{equation}
Similarly, for the radial and angular components of the stress-energy tensor one 
has

\begin{eqnarray}
t_{r}^{r} &=&
-\frac{\xi ^3 \left(r_{0}^2-2 \epsilon \right)^2 \left(r_{0}^2+4 \epsilon 
\right)}{8 r_{0}^6 \epsilon^3}\nonumber \\
&+&\frac{\xi ^2 \left(3 r_{0}^6-26r_{0}^2 \epsilon ^2+4 \epsilon ^3\right)}{16 
r_{0}^6 \epsilon^3}\nonumber \\
&+&\frac{\xi  \left(-15 r_{0}^6+4 r_{0}^4 \epsilon +102 r_{0}^2 \epsilon 
^2+8\epsilon^3\right)}{240 r_{0}^6 \epsilon^3}\nonumber \\
&+&\frac{22 r_{0}^6-20 r_{0}^4 \epsilon -107 r_{0}^2 \epsilon^2+58 
\epsilon^3}{3360 r_{0}^6\epsilon^3},
\end{eqnarray}

\begin{equation}
s_{r}^{r} =
\frac{17 r_{0}^6-74 r_{0}^4 \epsilon -20 r_{0}^2 \epsilon ^2+524 \epsilon 
^3}{6720 r_{0}^6 \epsilon ^3},
\end{equation}

\begin{equation}
v_{r}^{r} =
-\frac{-66r_{0}^6+228 r_{0}^4 \epsilon +223 r_{0}^2 \epsilon ^2-1490 \epsilon 
^3}{3360 r_{0}^6 \epsilon ^3},
\end{equation}

\begin{eqnarray}
t_{\theta}^{\theta} &=&
-\frac{\xi^3 \left(23 r_{0}^6-96 r_{0}^4 \epsilon +36 r_{0}^2 \epsilon^2+128 
\epsilon^3\right)}{16 r_{0}^6 \epsilon^3}\nonumber \\
&+&\frac{\xi^2 \left(87r_{0}^6-280 r_{0}^4 \epsilon -44 r_{0}^2 \epsilon^2-136 
\epsilon^3\right)}{32 r_{0}^6 \epsilon^3}\nonumber \\
&+&\frac{\xi  \left(-441 r_{0}^6+1328 r_{0}^4 \epsilon +460 r_{0}^2 
\epsilon^2+1016 \epsilon^3\right)}{480 r_{0}^6 \epsilon^3}\nonumber \\
&-&\frac{-283 r_{0}^6+833 r_{0}^4 \epsilon +382 r_{0}^2 \epsilon ^2+538 
\epsilon^3}{3360 r_{0}^6 \epsilon^3},
\end{eqnarray}

\begin{equation}
s_{\theta}^{\theta} =\frac{-143 r_{0}^6+224 r_{0}^4 \epsilon +382 r_{0}^2 
\epsilon ^2+1560 \epsilon ^3}{13440 r_{0}^6 \epsilon ^3}
\end{equation}
and
\begin{equation}
v_{\theta}^{\theta}=\frac{-95 r_{0}^6+175 r_{0}^4 \epsilon +416 r_{0}^2 \epsilon 
^2+106 \epsilon ^3}{1120 r_{0}^6 \epsilon ^3}.
\end{equation}
The simplicity of the metric is reflected in the simplicity of the resulting 
stress-energy tensors.
Analysis of the energy density and the radial tension of the spinor and vector 
fields indicates that the Morris-Thorne 
conditions cannot be satisfied. Specifically, the energy density is negative for 
$\varepsilon < 0.3084$ and $\varepsilon < 0.3153,$
for massive spinors and vectors, respectively. The radial pressure is always 
negative.
On the other hand,
the quantized massive scalar field has three regions in the parameter space in 
which the Morris-Thorne conditions are satisfied.
It should be noted however that neither minimal nor conformal couplings belong 
to these regions. 

\subsection{Wormhole supported by phantom energy}
\begin{figure}
 \centering
 \includegraphics[width=12cm]{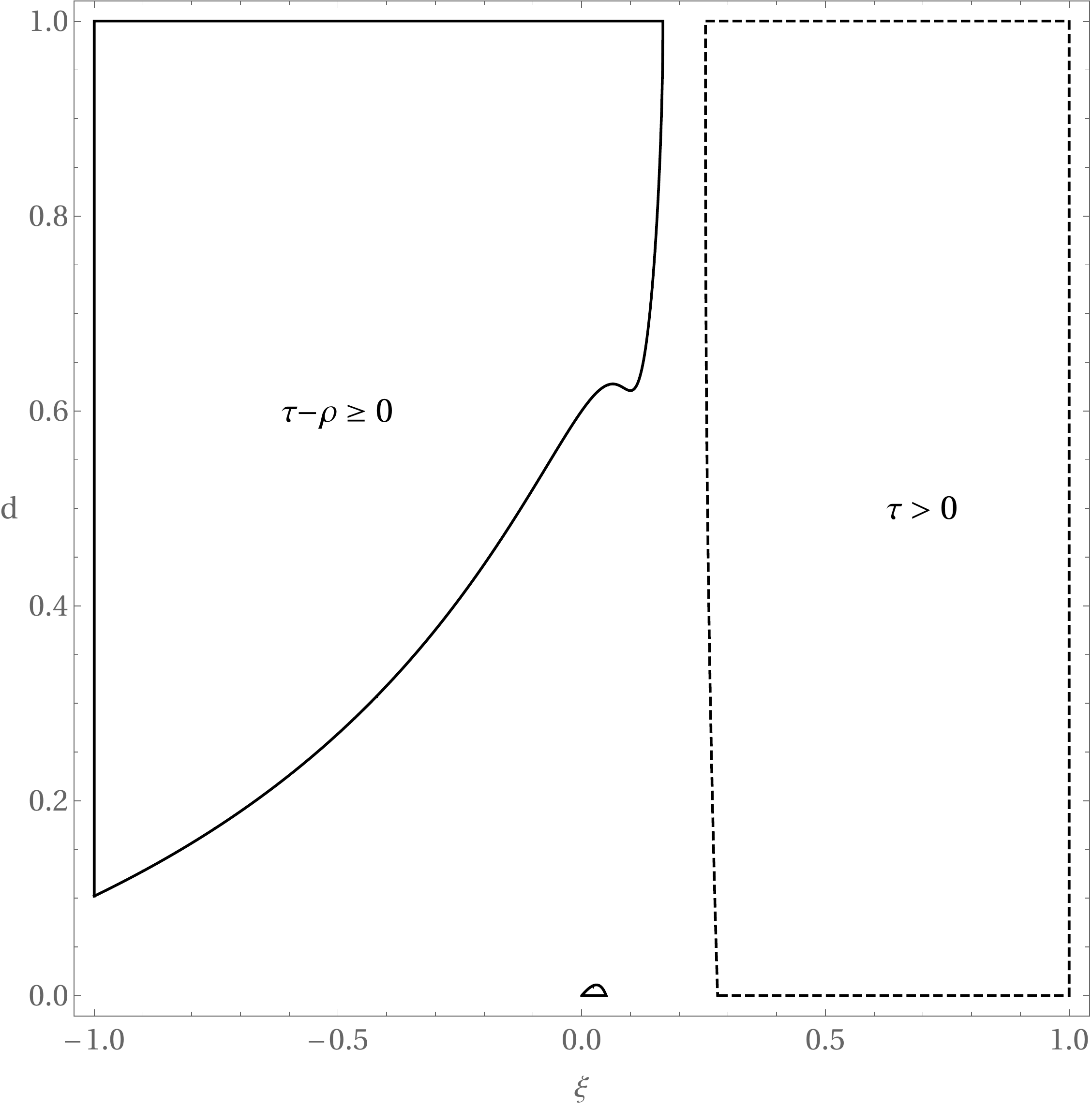}
 \caption{The regions of the rectangle $-1<\xi<1$, $0<d<1$ for which the 
exoticity
 condition  is satisfied (solid line).  The radial tension is positive within 
the 
 region bouded by a dashed line.  }
 \label{rys1}
\end{figure}

Finally, we shall analyze the  wormhole model  proposed by 
Zaslavskii~\cite{Oleg}. In this model the phantom energy
supports the wormhole described by the line element~(\ref{gen_el}) with
\begin{equation}
 \phi(r) = \frac{1}{2} \ln \frac{r_{1}}{r} \hspace{0.5cm} {\rm and} 
\hspace{0.5cm} b(r) = r_{0}+ d(r- r_{0})
\end{equation}
for $r_{0} \leq r \leq r_{b}$ and the Schwarzschild metric for $r>r_{b}.$ 
Here
$r_{1}$ is some constant that affects the normalization of time (our final 
results do
not depend on it) and $d<1.$ Let us observe that the wormhole is not 
asymptotically
flat. It can, however, be matched to the external Schwarzschild spacetime at 
some 
fixed radius.
The stress-energy tensor of the scalar, spinor and vector fields is given by 
Eq.~(\ref{strr})
with
\begin{eqnarray}
t_{t}^{t} &=& {\frac { \left( 9\,d-11 \right)  \left( 1+3\,d \right) 
^{2}\xi^{3}}{ 8 r_{0}^{6}}}-{\frac { \left( 1+3\,d \right)  \left( 
57\,d^{2}-84\,d+19 \right) \xi^{2}}{16 r_{0}^{6}}}\nonumber \\
&&
+{\frac {\left( 735\,d^{3}-1113\,d^{2}+277\,d+5 \right) 
\xi}{240\,r_{0}^{6}}}-{\frac 
{823\,d^{3}-1391\,d^{2}+537\,d-97}{3360\,r_{0}^{6}}},
\end{eqnarray}

\begin{equation}
s_{t}^{t} = \frac{786\, d^{3} -2519\, d^{2} + 2792\, d -899}{6720\, r_{0}^{6}},
\end{equation}

\begin{equation}
v_{t}^{t} = \frac{4013\,d^{3} -11325\,d^{2} + 11283 \,d - 
3587}{3360\,r_{0}^{6}},
\end{equation}
\begin{eqnarray}
t_{r}^{r} &=& \frac { \left( 3\,d-5 \right)  \left( 1+3\,d \right) 
^{2}\xi^{3}}{8 r_{0}^{6}}-\frac { \left( 1+3\,d\right)  \left( 15\,d^{2}-24\,d+1 
\right) \xi^{2}}{ 16 r_{0}^{6}}\nonumber \\
	&&
+\frac { \left( 63\,d^{3}-97\,d^{2}+5\,d-3\right) \xi}{80\, r_{0}^{6} }-\frac 
{263\,d^{3}-495\,d^{2}+201\,d-97}{3360\,r_{0}^{6}},
\end{eqnarray}
	
	\begin{equation}
s_{r}^{r} = \frac{-208\,d^{3} +631 \,d^{2} -638\,d +375}{6720\,r_{0}^{6}},
	\end{equation}
	
\begin{equation}
v_{r}^{r} = \frac{-915\,d^{3} +2227\,d^{2} -1709\,d + 781}{3360 \,r_{0}^{6}},
\end{equation}

\begin{eqnarray}
	t_{\theta}^{\theta} &=& {\frac { \left( 27\,d-19 \right)  \left( 1+3\,d 
\right) ^{2}\xi^{3}}{ 16 r_{0}^{6}}}-{\frac { \left( 1+3\,d \right)  \left( 
153\,d^{2}-158\,d+37 \right) \xi^{2}}{32 r_{0}^{6}}}
	\nonumber \\
	&&
	+{\frac { \left( 2037\,d^{3}-2391\,d^{2}+655\,d+83 \right) 
\xi}{480\,r_{0}^{6}}}-{\frac 
{1411\,d^{3}-2273\,d^{2}+1265\,d-147}{3360\,r_{0}^{6}}},
\end{eqnarray}

\begin{equation}
s_{\theta}^{\theta} = - \frac{1437\,d^{3} -4419 \,d^{2} +4811\,d 
-1225}{13440\,r_{0}^{6}}
\end{equation}
and
\begin{equation}
v_{\theta}^{\theta} = -\frac{2791\,d^{3} -5573\, d^{2} + 
3949\,d-399}{3360\,r_{0}^{6}}.
\end{equation}
Let us consider the stress-energy tensor of the quantized spinor and vector 
fields first. 
The energy density is positive for $d<0.5428$ for massive spinors and for 
$d<0.6139$
for vectors. The radial tension is always negative and the second Morris-Thorne
condition is violated. Thus there are no configurations that sustain 
a Zaslavskii wormhole.
The parameter space  for the massive scalar field is two-dimensional.
Simple analysis shows that there are no configurations that satisfy the  
Morris-Thorne
conditions as can easily be inferred form Fig.~\ref{rys1}. (The regions of 
the
parameter space defined by the Morris-Thorne conditions are disjoint.)

\section{Final remarks}
\label{fin}
In this paper, we have attempted to  answer the question posed in 
Ref.~\cite{TaylorPaul} of
whether the stress-energy tensor of the quantized  massive fields has the 
appropriate 
form  to support the traversable wormhole. The scalar case has been considered 
earlier 
by Taylor, Hiscock and Anderson (cases A - E). Our results are in concord with 
the previous 
studies and indicate that the quantized massive scalar, spinor and vector fields 
do not 
satisfy simultaneously the Morris-Thorne conditions. Here we adopted a somewhat 
conservative 
point of view and treated as physical only the minimal and conformal curvature 
couplings. 
The reported calculations indicate serious problems one encounters in 
the maintenance of a
traversable wormhole.  Indeed, regardless of the spin, the quantum fields tend 
to 
destroy a wormhole and this may suggest that even if the fields with the 
desired properties 
exist, the net effect would be negative. It is simply because the fields that do 
not satisfy
the Morris-Thorne conditions  would also be present there. We have borrowed the 
last argument 
from Ref.~\cite{TaylorPaul}. This, of course, does not exclude the possibility 
of the existence 
of the wormholes  as self-consistent solutions of the semiclassical 
Einstein field 
equations
\begin{equation}
 G_{ab} + \alpha I_{ab} + \beta J_{ab}= 8\pi \langle  T_{ab} \rangle,
\end{equation} 
with the quadratic terms $I_{ab}$ and $J_{ab},$ the coupling constants (usually 
set to zero) 
$\alpha,\,\beta$ and with the source term given by the stress-energy tensor of 
the quantized 
fields functionally depending on the metric. In  this case, however, we are 
looking for 
the particular solution (or a class of solutions that satisfy prescribed 
requirements) of 
the system of the  differential equations. For the stress-energy tensor 
constructed within 
the framework of the Schwinger-DeWitt approximation, the problem 
reduces to a complicated system of  differential equations, which must be 
studied
numerically.  To be more precise, let us consider the line element describing 
a wormhole in the form
\begin{equation}
 ds^{2} = -f(l) dt^{2} + dl^{2} + r^{2}(l) (d\theta^{2}+\sin^{2} \theta 
d\phi^{2}),
\end{equation}
where $l$ is the proper distance from the throat and $f$ and $r$ are two unknown 
functions. 
The resulting $(ll)$-component of the semiclassical Einstein field equations is 
of the fifth order,
whereas the remaining ones are of the sixth order. These equations, with  
carefully chosen 
initial conditions,  can be analyzed numerically in the spirit of 
Ref.~\cite{PopovSushkov}.

The results presented in this paper raise the interesting question of whether
obstacles one encounters in constructing and maintaining the traversable 
wormhole are specific to 
the four-dimensional world or generic. The techniques described here 
may shed light on this problem.  Indeed, since the Schwinger-DeWitt approach 
can be extended to other dimensions~\cite{kocio_dim_a,kocio_dim_b,kocio_dim_c} 
(see also Refs.~\cite{Lemos,br1,br2}) one can obtain valuable information 
regarding 
the general nature of the traversable wormholes. Indeed, let us concentrate 
on the quantized massive scalar field. For $D=4$ and $D=5$ the effective action 
$W^{(1)}$ is constructed from the coefficient $[a_{3}]$ whereas for $D=6$ and 
$D=7$ the coefficient $[a_{4}]$ is needed. The stress-energy tensor can be 
obtained in a standard way, i.e., by taking the functional derivative of the 
action with respect to the metric and  the resulting general formulas  can be 
used in the spacetime of the higher-dimensional wormhole. As the computational 
complexity of the problem rapidly grows with the dimension $D$, the 
(extremely complicated and time-consuming) calculations are practically limited 
to
$D\leq 9.$ All these problems are under active investigation and we intend to 
return to them in the future.

\begin{acknowledgments}
Discussions with Pawe{\l} Sadurski are gratefully acknowledged.
J.M. was partially supported by the Polish National Science Centre 
grant no. DEC-2014/15/B/ST2/00089.
\end{acknowledgments}


\begin{thebibliography}{33}
	\expandafter\ifx\csname natexlab\endcsname\relax\def\natexlab#1{#1}\fi
	\expandafter\ifx\csname bibnamefont\endcsname\relax
	\def\bibnamefont#1{#1}\fi
	\expandafter\ifx\csname bibfnamefont\endcsname\relax
	\def\bibfnamefont#1{#1}\fi
	\expandafter\ifx\csname citenamefont\endcsname\relax
	\def\citenamefont#1{#1}\fi
	\expandafter\ifx\csname url\endcsname\relax
	\def\url#1{\texttt{#1}}\fi
	\expandafter\ifx\csname urlprefix\endcsname\relax\def\urlprefix{URL }\fi
	\providecommand{\bibinfo}[2]{#2}
	\providecommand{\eprint}[2][]{\url{#2}}
	
	\bibitem[{\citenamefont{Morris and Thorne}(1988)}]{Morris1}
	\bibinfo{author}{\bibfnamefont{M.~S.} \bibnamefont{Morris}} \bibnamefont{and}
	\bibinfo{author}{\bibfnamefont{K.~S.} \bibnamefont{Thorne}},
	\bibinfo{journal}{Am. J. Phys.} \textbf{\bibinfo{volume}{56}},
	\bibinfo{pages}{395} (\bibinfo{year}{1988}).
	
	\bibitem[{\citenamefont{Morris et~al.}(1988)\citenamefont{Morris, Thorne, and
			Yurtsever}}]{Morris2}
	\bibinfo{author}{\bibfnamefont{M.~S.} \bibnamefont{Morris}},
	\bibinfo{author}{\bibfnamefont{K.~S.} \bibnamefont{Thorne}},
	\bibnamefont{and}
	\bibinfo{author}{\bibfnamefont{U.}~\bibnamefont{Yurtsever}},
	\bibinfo{journal}{Phys. Rev. Lett.} \textbf{\bibinfo{volume}{61}},
	\bibinfo{pages}{1446} (\bibinfo{year}{1988}).
	
	\bibitem[{\citenamefont{Visser}(1995)}]{VisserBook}
	\bibinfo{author}{\bibfnamefont{M.}~\bibnamefont{Visser}},
	\emph{\bibinfo{title}{{Lorentzian wormholes: From Einstein to Hawking}}}
	(\bibinfo{publisher}{American Institute of Physics},
	\bibinfo{address}{Woodbury, NY}, \bibinfo{year}{1995}).
	
	\bibitem[{\citenamefont{Lobo}(2016)}]{Lobo}
	\bibinfo{author}{\bibfnamefont{F.~S.~N.} \bibnamefont{Lobo}},
	\bibinfo{journal}{Int. J. Mod. Phys.} \textbf{\bibinfo{volume}{D25}},
	\bibinfo{pages}{1630017} (\bibinfo{year}{2016}).
	
	\bibitem[{\citenamefont{Bronnikov and Rubin}(2012)}]{KirillBook}
	\bibinfo{author}{\bibfnamefont{K.~A.} \bibnamefont{Bronnikov}}
	\bibnamefont{and} \bibinfo{author}{\bibfnamefont{S.~G.} \bibnamefont{Rubin}},
	\emph{\bibinfo{title}{{Black Holes, Cosmology and Extra Dimensions}}}
	(\bibinfo{publisher}{World Scientific}, \bibinfo{address}{Singapore},
	\bibinfo{year}{2012}).
	
	\bibitem[{\citenamefont{Ayón-Beato et~al.}(2016)\citenamefont{Ayón-Beato,Canfora, and Zanelli}}]{ayon}
        \bibinfo{author}{\bibfnamefont{E.}~\bibnamefont{Ayon-Beato}},
        \bibinfo{author}{\bibfnamefont{F.}~\bibnamefont{Canfora}}, \bibnamefont{and}
        \bibinfo{author}{\bibfnamefont{J.}~\bibnamefont{Zanelli}},
        \bibinfo{journal}{Physics Letters B} \textbf{\bibinfo{volume}{752}},
        \bibinfo{pages}{201 } (\bibinfo{year}{2016}).
	
	\bibitem[{\citenamefont{Taylor et~al.}(1997)\citenamefont{Taylor, Hiscock, and
			Anderson}}]{TaylorPaul}
	\bibinfo{author}{\bibfnamefont{B.~E.} \bibnamefont{Taylor}},
	\bibinfo{author}{\bibfnamefont{W.~A.} \bibnamefont{Hiscock}},
	\bibnamefont{and} \bibinfo{author}{\bibfnamefont{P.~R.}
		\bibnamefont{Anderson}}, \bibinfo{journal}{Phys. Rev.}
	\textbf{\bibinfo{volume}{D55}}, \bibinfo{pages}{6116} (\bibinfo{year}{1997}).
	
	\bibitem[{\citenamefont{Anderson et~al.}(1995)\citenamefont{Anderson, Hiscock,
			and Samuel}}]{Paul}
	\bibinfo{author}{\bibfnamefont{P.~R.} \bibnamefont{Anderson}},
	\bibinfo{author}{\bibfnamefont{W.~A.} \bibnamefont{Hiscock}},
	\bibnamefont{and} \bibinfo{author}{\bibfnamefont{D.~A.}
		\bibnamefont{Samuel}}, \bibinfo{journal}{Phys. Rev.}
	\textbf{\bibinfo{volume}{D51}}, \bibinfo{pages}{4337} (\bibinfo{year}{1995}).
	
	\bibitem[{\citenamefont{Carlson et~al.}(2010)\citenamefont{Carlson, Anderson,
			Fabbri, Fagnocchi, Hirsch, and Klyap}}]{Carlson}
	\bibinfo{author}{\bibfnamefont{E.~D.} \bibnamefont{Carlson}},
	\bibinfo{author}{\bibfnamefont{P.~R.} \bibnamefont{Anderson}},
	\bibinfo{author}{\bibfnamefont{A.}~\bibnamefont{Fabbri}},
	\bibinfo{author}{\bibfnamefont{S.}~\bibnamefont{Fagnocchi}},
	\bibinfo{author}{\bibfnamefont{W.~H.} \bibnamefont{Hirsch}},
	\bibnamefont{and} \bibinfo{author}{\bibfnamefont{S.}~\bibnamefont{Klyap}},
	\bibinfo{journal}{Phys. Rev.} \textbf{\bibinfo{volume}{D82}},
	\bibinfo{pages}{124070} (\bibinfo{year}{2010}).
	
	\bibitem[{\citenamefont{Khatsymovsky}(1997)}]{Khatsymovsky}
	\bibinfo{author}{\bibfnamefont{V.}~\bibnamefont{Khatsymovsky}},
	\bibinfo{journal}{Phys. Lett.} \textbf{\bibinfo{volume}{B399}},
	\bibinfo{pages}{215} (\bibinfo{year}{1997}).
	
	\bibitem[{\citenamefont{Popov}(2014)}]{Popov2014}
	\bibinfo{author}{\bibfnamefont{A.~A.} \bibnamefont{Popov}},
	\bibinfo{journal}{Grav. Cosmol.} \textbf{\bibinfo{volume}{20}},
	\bibinfo{pages}{203} (\bibinfo{year}{2014}).
	
	\bibitem[{\citenamefont{Bezerra et~al.}(2010)\citenamefont{Bezerra, Bezerra~de
			Mello, Khusnutdinov, and Sushkov}}]{Bezerra}
	\bibinfo{author}{\bibfnamefont{V.~B.} \bibnamefont{Bezerra}},
	\bibinfo{author}{\bibfnamefont{E.~R.} \bibnamefont{Bezerra~de Mello}},
	\bibinfo{author}{\bibfnamefont{N.~R.} \bibnamefont{Khusnutdinov}},
	\bibnamefont{and} \bibinfo{author}{\bibfnamefont{S.~V.}
		\bibnamefont{Sushkov}}, \bibinfo{journal}{Phys. Rev.}
	\textbf{\bibinfo{volume}{D81}}, \bibinfo{pages}{084034}
	(\bibinfo{year}{2010}.
	
	\bibitem[{\citenamefont{Popov and Sushkov}(2001)}]{PopovS2}
	\bibinfo{author}{\bibfnamefont{A.~A.} \bibnamefont{Popov}} \bibnamefont{and}
	\bibinfo{author}{\bibfnamefont{S.~V.} \bibnamefont{Sushkov}},
	\bibinfo{journal}{Phys. Rev.} \textbf{\bibinfo{volume}{D63}},
	\bibinfo{pages}{044017} (\bibinfo{year}{2001}).
	
	\bibitem[{\citenamefont{Popov}(2001)}]{Popov2001}
	\bibinfo{author}{\bibfnamefont{A.~A.} \bibnamefont{Popov}},
	\bibinfo{journal}{Phys. Rev.} \textbf{\bibinfo{volume}{D64}},
	\bibinfo{pages}{104005} (\bibinfo{year}{2001}).
	
	\bibitem[{\citenamefont{Khusnutdinov}(2003)}]{Khu}
	\bibinfo{author}{\bibfnamefont{N.~R.} \bibnamefont{Khusnutdinov}},
	\bibinfo{journal}{Phys. Rev.} \textbf{\bibinfo{volume}{D67}},
	\bibinfo{pages}{124020} (\bibinfo{year}{2003}).
	
	\bibitem[{\citenamefont{Garattini and Lobo}(2007)}]{Garattini}
	\bibinfo{author}{\bibfnamefont{R.}~\bibnamefont{Garattini}} \bibnamefont{and}
	\bibinfo{author}{\bibfnamefont{F.~S.~N.} \bibnamefont{Lobo}},
	\bibinfo{journal}{Class. Quant. Grav.} \textbf{\bibinfo{volume}{24}},
	\bibinfo{pages}{2401} (\bibinfo{year}{2007}).
	
	\bibitem[{\citenamefont{Hochberg et~al.}(1997)\citenamefont{Hochberg, Popov,
			and Sushkov}}]{PopovSushkov}
	\bibinfo{author}{\bibfnamefont{D.}~\bibnamefont{Hochberg}},
	\bibinfo{author}{\bibfnamefont{A.}~\bibnamefont{Popov}}, \bibnamefont{and}
	\bibinfo{author}{\bibfnamefont{S.~V.} \bibnamefont{Sushkov}},
	\bibinfo{journal}{Phys. Rev. Lett.} \textbf{\bibinfo{volume}{78}},
	\bibinfo{pages}{2050} (\bibinfo{year}{1997}).
	
	\bibitem[{\citenamefont{DeWitt}(1965)}]{Bryce1}
	\bibinfo{author}{\bibfnamefont{B.~S.} \bibnamefont{DeWitt}},
	\emph{\bibinfo{title}{Dynamical Theory of groups and fields}}
	(\bibinfo{publisher}{Gordon and Breach}, \bibinfo{address}{New York},
	\bibinfo{year}{1965}).
	
	\bibitem[{\citenamefont{Avramidi}(2000)}]{IvanBook}
	\bibinfo{author}{\bibfnamefont{I.~G.} \bibnamefont{Avramidi}},
	\emph{\bibinfo{title}{Heat kernel and quantum gravity}}
	(\bibinfo{publisher}{Springer-Verlag}, \bibinfo{address}{Berlin},
	\bibinfo{year}{2000}).
	
	\bibitem[{\citenamefont{Frolov and Zelnikov}(1984)}]{Frolov}
	\bibinfo{author}{\bibfnamefont{V.~P.} \bibnamefont{Frolov}} \bibnamefont{and}
	\bibinfo{author}{\bibfnamefont{A.}~\bibnamefont{Zelnikov}},
	\bibinfo{journal}{Phys.Rev.} \textbf{\bibinfo{volume}{D29}},
	\bibinfo{pages}{1057} (\bibinfo{year}{1984}).
	
	\bibitem[{\citenamefont{Matyjasek}(2000)}]{Kocio1}
	\bibinfo{author}{\bibfnamefont{J.}~\bibnamefont{Matyjasek}},
	\bibinfo{journal}{Phys. Rev.} \textbf{\bibinfo{volume}{D61}},
	\bibinfo{pages}{124019} (\bibinfo{year}{2000}).
	
	\bibitem[{\citenamefont{Matyjasek}(2001)}]{Kocio2}
	\bibinfo{author}{\bibfnamefont{J.}~\bibnamefont{Matyjasek}},
	\bibinfo{journal}{Phys. Rev.} \textbf{\bibinfo{volume}{D63}},
	\bibinfo{pages}{084004} (\bibinfo{year}{2001}).
	
	\bibitem[{\citenamefont{Taylor et~al.}(2000)\citenamefont{Taylor, Hiscock, and
			Anderson}}]{Paul1}
	\bibinfo{author}{\bibfnamefont{B.~E.} \bibnamefont{Taylor}},
	\bibinfo{author}{\bibfnamefont{W.~A.} \bibnamefont{Hiscock}},
	\bibnamefont{and} \bibinfo{author}{\bibfnamefont{P.~R.}
		\bibnamefont{Anderson}}, \bibinfo{journal}{Phys. Rev.}
	\textbf{\bibinfo{volume}{D61}}, \bibinfo{pages}{084021}
	(\bibinfo{year}{2000}).
	
	\bibitem[{\citenamefont{Matyjasek and Tryniecki}(2009)}]{next}
	\bibinfo{author}{\bibfnamefont{J.}~\bibnamefont{Matyjasek}} \bibnamefont{and}
	\bibinfo{author}{\bibfnamefont{D.}~\bibnamefont{Tryniecki}},
	\bibinfo{journal}{Phys. Rev.} \textbf{\bibinfo{volume}{D79}},
	\bibinfo{pages}{084017} (\bibinfo{year}{2009}).
	
	\bibitem[{\citenamefont{Matyjasek}(2016)}]{kocio_dim_a}
	\bibinfo{author}{\bibfnamefont{J.}~\bibnamefont{Matyjasek}},
	\bibinfo{journal}{Phys. Rev.} \textbf{\bibinfo{volume}{D94}},
	\bibinfo{pages}{084048} (\bibinfo{year}{2016}).
	
	\bibitem[{\citenamefont{Matyjasek and
			Sadurski}(2015{\natexlab{a}})}]{kocio_dim_b}
	\bibinfo{author}{\bibfnamefont{J.}~\bibnamefont{Matyjasek}} \bibnamefont{and}
	\bibinfo{author}{\bibfnamefont{P.}~\bibnamefont{Sadurski}},
	\bibinfo{journal}{Phys. Rev.} \textbf{\bibinfo{volume}{D92}},
	\bibinfo{pages}{044023} (\bibinfo{year}{2015}{\natexlab{a}}).
	
	\bibitem[{\citenamefont{Matyjasek and
			Sadurski}(2015{\natexlab{b}})}]{kocio_dim_c}
	\bibinfo{author}{\bibfnamefont{J.}~\bibnamefont{Matyjasek}} \bibnamefont{and}
	\bibinfo{author}{\bibfnamefont{P.}~\bibnamefont{Sadurski}},
	\bibinfo{journal}{Phys. Rev.} \textbf{\bibinfo{volume}{D91}},
	\bibinfo{pages}{044027} (\bibinfo{year}{2015}{\natexlab{b}}).
	
	\bibitem[{\citenamefont{Zaslavskii}(2005)}]{Oleg}
	\bibinfo{author}{\bibfnamefont{O.~B.} \bibnamefont{Zaslavskii}},
	\bibinfo{journal}{Phys. Rev.} \textbf{\bibinfo{volume}{D72}},
	\bibinfo{pages}{061303} (\bibinfo{year}{2005}.
	
	\bibitem[{\citenamefont{Vermaseren}(2000)}]{Jos1}
	\bibinfo{author}{\bibfnamefont{J.~A.~M.} \bibnamefont{Vermaseren}}
	(\bibinfo{year}{2000}), \eprint{arXiv:math-ph/0010025}.
	
	\bibitem[{\citenamefont{Ruijl et~al.}(2017)\citenamefont{Ruijl, Ueda, and
			Vermaseren}}]{Jos2}
	\bibinfo{author}{\bibfnamefont{B.}~\bibnamefont{Ruijl}},
	\bibinfo{author}{\bibfnamefont{T.}~\bibnamefont{Ueda}}, \bibnamefont{and}
	\bibinfo{author}{\bibfnamefont{J.}~\bibnamefont{Vermaseren}}
	(\bibinfo{year}{2017}), \eprint{arXiv:1707.06453}.
	
	\bibitem[{\citenamefont{Thompson and Lemos}(2009)}]{Lemos}
	\bibinfo{author}{\bibfnamefont{R.~T.} \bibnamefont{Thompson}} \bibnamefont{and}
	\bibinfo{author}{\bibfnamefont{J.~P.~S.} \bibnamefont{Lemos}},
	\bibinfo{journal}{Phys. Rev.} \textbf{\bibinfo{volume}{D80}},
	\bibinfo{pages}{064017} (\bibinfo{year}{2009}).
	
	\bibitem[{\citenamefont{Taylor and Breen}(2016)}]{br1}
	\bibinfo{author}{\bibfnamefont{P.}~\bibnamefont{Taylor}} \bibnamefont{and}
	\bibinfo{author}{\bibfnamefont{C.}~\bibnamefont{Breen}},
	\bibinfo{journal}{Phys. Rev.} \textbf{\bibinfo{volume}{D94}},
	\bibinfo{pages}{125024} (\bibinfo{year}{2016}).
	
	\bibitem[{\citenamefont{Taylor and Breen}(2017)}]{br2}
	\bibinfo{author}{\bibfnamefont{P.}~\bibnamefont{Taylor}} \bibnamefont{and}
	\bibinfo{author}{\bibfnamefont{C.}~\bibnamefont{Breen}}
	(\bibinfo{year}{2017}), \eprint{arXiv:1709.00316}.
	
\end{thebibliography}

\end{document}